\documentclass[prb,a4,twocolumn,amsmath,amssymb]{revtex4}
\usepackage{graphicx}
\usepackage{dcolumn}
\usepackage{bm}

\begin{document}

\title{A versatile dual spot laser scanning confocal microscopy system\\for advanced fluorescence correlation spectroscopy analysis in living cell}

\author{Patrick Ferrand}
\email{patrick.ferrand@fresnel.fr}
\affiliation{Aix-Marseille Universit\'e, Institut Fresnel, Campus de St J\'er\^ome, F-13013 Marseille, France}
\affiliation{Ecole Centrale Marseille, Institut Fresnel, Campus de St J\'er\^ome, F-13013 Marseille, France}
\affiliation{CNRS, Institut Fresnel, Campus de St J\'er\^ome, F-13013 Marseille, France}
\author{Martina Pianta}
\affiliation{Aix-Marseille Universit\'e, Institut Fresnel, Campus de St J\'er\^ome, F-13013 Marseille, France}
\affiliation{Ecole Centrale Marseille, Institut Fresnel, Campus de St J\'er\^ome, F-13013 Marseille, France}
\affiliation{CNRS, Institut Fresnel, Campus de St J\'er\^ome, F-13013 Marseille, France}
\author{Alla Kress}
\affiliation{Aix-Marseille Universit\'e, Institut Fresnel, Campus de St J\'er\^ome, F-13013 Marseille, France}
\affiliation{Ecole Centrale Marseille, Institut Fresnel, Campus de St J\'er\^ome, F-13013 Marseille, France}
\affiliation{CNRS, Institut Fresnel, Campus de St J\'er\^ome, F-13013 Marseille, France}
\author{Alexandre Aillaud}
\affiliation{Aix-Marseille Universit\'e, Institut Fresnel, Campus de St J\'er\^ome, F-13013 Marseille, France}
\affiliation{Ecole Centrale Marseille, Institut Fresnel, Campus de St J\'er\^ome, F-13013 Marseille, France}
\affiliation{CNRS, Institut Fresnel, Campus de St J\'er\^ome, F-13013 Marseille, France}
\author{Herv\'e Rigneault}
\affiliation{Aix-Marseille Universit\'e, Institut Fresnel, Campus de St J\'er\^ome, F-13013 Marseille, France}
\affiliation{Ecole Centrale Marseille, Institut Fresnel, Campus de St J\'er\^ome, F-13013 Marseille, France}
\affiliation{CNRS, Institut Fresnel, Campus de St J\'er\^ome, F-13013 Marseille, France}
\author{Didier Marguet}
\affiliation{Aix-Marseille Universit\'e, CIML, Parc scientifique de Luminy, Case 906, F-13288 Marseille, France}
\affiliation{INSERM, CIML, Parc scientifique de Luminy, Case 906, F-13288 Marseille, France}
\affiliation{CNRS, CIML, Parc scientifique de Luminy, Case 906, F-13288 Marseille, France}

\date{\today}

\begin{abstract}
A fluorescence correlation spectroscopy (FCS) system based on two independent measurement volumes is presented. The optical setup and data acquisition hardware are detailed, as well as a complete protocol to control the location, size and shape of the measurement volumes. A method that allows to monitor independently the excitation and collection efficiency distribution is proposed. Finally, a few examples of measurements that exploit the two spots in static and/or scanning schemes, are reported.
\end{abstract}

\pacs{Valid PACS appear here}

\maketitle

\section{Introduction}

Fluorescence correlation spectroscopy (FCS) is a technique based on the analysis of the fluctuations of fluorescence that permits to quantify a wide range of phenomena such as photophysical, photochemical, interaction, diffusion and transport properties of fluorescently labeled molecules.\cite{Webb2001} Benefitting from the dramatic progress of bright fluorescent molecules and high sensitivity detectors, this technique can now be performed within the confocal volume of high numerical aperture microscope objectives. The spatial resolution, the intrinsic steady-state regime, as well as the capability to record a huge number of events make it now a routine tool for cell biology.\cite{Schwille2001}

However, the original FCS technique is limited by the fact the dynamic system under study is exclusively analyzed in terms of temporal fluctuations at a single location, at a spatial scale related to the measurement volume, while a much more complete description would require investigations in the spatio-temporal domain. As a consequence, various phenomena of similar timescale taking place within the same specimen cannot be efficiently discriminated without a strong a priori knowledge of the system. For instance, FCS measurements of translational diffusion performed under photobleaching conditions may actually monitor the survival time of the fluorophores instead of their diffusion time.\cite{Eggeling1998}

In order to overcome this lack of spatial information, several FCS-based techniques have been proposed. The analysis can be performed by cross-correlating signals recorded at two distant locations; this approach suits the best to transport phenomena\cite{Brinkmeier1999} but it can also successfully be extended to the measurement of absolute diffusion, although the time gating that is required in order to overcome the crosstalk between the two overlapping volumes makes it more difficult to implement.\cite{Dertinger2007} The two observation volumes can also be simply realized by splitting the detected fluorescence onto two shifted detectors,\cite{Jaffiol2006} but this causes the two resulting measurement volumes to be slightly deformed. A more versatile scheme has been proposed using an array detector, but it is still limited by a finite readout speed.\cite{Burkhardt2006} Another approach allowing to achieve a better spatial description of a system is to perform FCS measurements at various spatial scales, by changing the observation volumes.\cite{Wawrezinieck2005} It has been shown that this approach allows to discriminate between possible diffusion regimes of molecular species in cell membranes and even to quantify submicron structures.\cite{Lenne2006}

Another class of methods derived from FCS consists in scanning the observation spot in a repetitive fashion in the sample, and therefore collecting sequentially fluorescence information from many locations, which improves the statistical accuracy of the measurement and reduces photobleaching.\cite{Weissman1973} Line or circle scan are used in so-called scanning FCS (sFCS) experiments, while image scans produce image frames that are analyzed using Image correlation spectroscopy (ICS) techniques. Although sFCS and ICS are based on the same principle, they address different time scales. sFCS can analyze dynamics in a wide range down to a fraction of a millisecond.\cite{Berland1996} It was shown recently that this method allows to measure absolute diffusion coefficient in biological systems,\cite{Petrasek2008a} and even to discriminate different dynamic processes of comparable timescale.\cite{Petrasek2008b} Except raster ICS (RICS), which allows to analyze fast dynamics by exploiting the time taken by the raster scan,\cite{Digman2005a} ICS and its variants\cite{Kolin2007} suit better to slowly diffusing systems. All these methods can be in principle easily implemented on any laser scanning confocal microscopy system.

In this article, we describe a FCS system that is based on two custom laser scanning confocal microscopes sharing the same objective, and creating within the specimen two fully independent diffraction-limited measurement spots. Each spot can operate either in a static mode at an arbitrary location or in a scanning mode, along an arbitrary periodic trajectory that can be one or several image frames, circles, lines, or any other closed loop, while the signals of fluorescence are recorded by two dedicated confocal detection channels. The instrument has been designed in order to offer a high level of flexibility. The operating modes include dual spot twin measurements (i.e., simultaneous measurements performed in the same mode at two locations), dual spot cross analysis, as well as hybrid measurements that assign a different measurement mode to each spot.

The article is organized as follows. Section~\ref{Sec:Setup} details the optical setup and hardware. Section~\ref{Sec:Theory} describes briefly the theoretical background of confocal microscopy and FCS measurements. Section~\ref{Sec:Calibrations} addresses the protocol of calibration of the system. Finally, section~\ref{Sec:Examples} reports examples of measurements exploiting the versatility and the dual spot nature of the system.

\section{Description of the setup}
\label{Sec:Setup}

\subsection{Optical bench}

The two observation volumes in our system are produced by using two identical custom laser scanning confocal microscopy systems sharing the same objective. Although this is likely the most complex technical approach for generating two measurement spots, we believe that it provides the highest level of versatility without sacrificing the optical quality of the spots.

\begin{figure}
\includegraphics{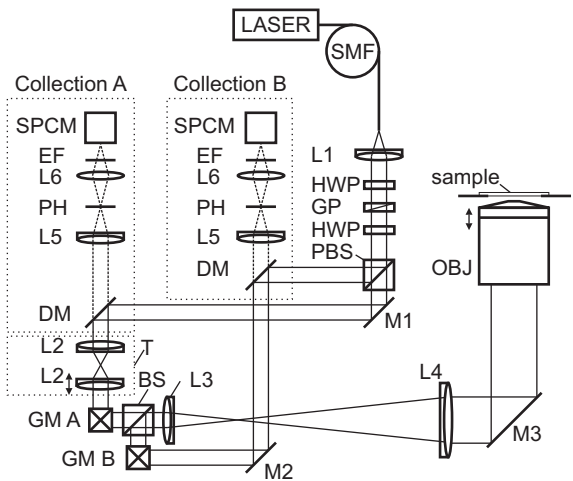}
\caption{\label{Fig:setup}Schematics of the dual spot FCS system. Abbreviations: SMF, single mode fiber; L1-L6, achromatic doublets (L1 = 25 mm, L2 = 30 mm, L3 = 100 mm, L4 = 300 mm, L5 = 150 mm, L6 = 50 mm); HWP, half-wave plate; GP, Glan polarizer; PBS, polarizing beamsplitter; M1-M3, mirrors; DM, dichroic mirror; T, telescope; GM A (resp., B), galvanomtric mirror set for channel A (resp., B); OBJ, water immersion microscope objective; PH, pinhole; EF, emission filter; SPCM: single photon counting module.}
\end{figure}

The optical setup is presented in Fig.~\ref{Fig:setup}. Emission from a continuous wave 491-nm diode-pumped solid state laser (Calypso, Cobolt) is coupled into a single mode fiber. The purpose of the single mode fiber is to allow a convenient coupling for other laser sources for future development of the system. Light exiting the fiber is collimated by an achromatic doublet and attenuated using a half-wave plate placed in front of a Glan polarizer. The beam is divided into two excitation arms by a polarizing beam splitter. A second half-wave plate located before the beam splitter controls the power ratio between the two arms. The two excitation beams follow paths of equal distance. They are first reflected by a dichroic mirror (XF2037-500DRLP, Omega Optical) and then sent on a couple of galvanometric scanning mirrors (6200H, Cambridge Technology). The two excitation beams are combined by a non-polarizing beam splitter and introduced into the side port of the microscope stand (Axiovert 200M, Carl Zeiss). A scanning telescope images the scanners with a magnification of 3 onto the rear aperture of the infinity corrected water immersion microscope objective (C-Apochromat $40\times$, focal length 4.1~mm, NA = 1.2, UV-VIS-NIR, Carl Zeiss). Two independent excitation spots, that we denote A and B, are created within the sample, each one at a location controlled by its dedicated scanning system.

 Fluorescence emitted at each spot location is collected by the same objective, follows the same path back through its corresponding scanner, in a so-called descanned scheme, and is sent through the dichroic mirror of the corresponding channel. The two detection benches are identically constituted of a 75-$\mu$m diameter pinhole (i.e., 1.2 Airy units with our total magnification of 120) placed at the focus of a tubelens. Another lens images (with a magnification of unity) the pinhole onto the 175-$\mu$m diameter active area of a single photon counting module (SPCM-AQR-14, PerkinElmer Optoelectronics).

This setup creates two identical excitation spots with their dedicated confocal detection, while the descanned detection scheme keeps the conjugation unaffected when the spots are moved. As illustrated in Fig.~\ref{Fig:setup}, an additional telescope (denoted T) of magnification $-1$ has been inserted on channel~A. By displacing one lens with respect to the other along the optical axis, it is possible to modify slightly the divergence of excitation beam A, and therefore to change the plane in which spot A is focused. Note that the conjugation between the spot and the pinhole is not affected because the divergence induced on the excitation is compensated on the detection beam. A procedure for calibrating the axial displacement of the focus and assessing the spot quality in this configuration will be presented in section~\ref{Sec:Calibrations}.

\subsection{Data acquisition}

\begin{figure}
\includegraphics{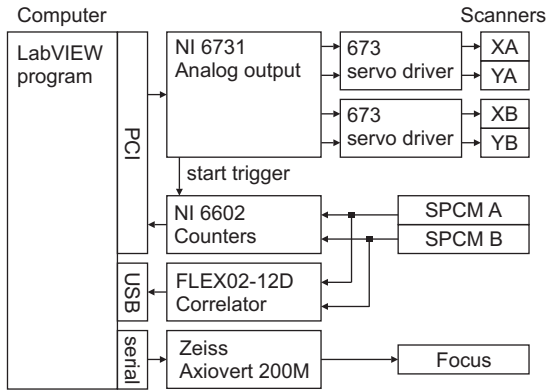}
\caption{\label{Fig:data_acquisition}Diagram of the acquisition hardware.}
\end{figure}

A diagram of the data acquisition hardware is presented in Fig.~\ref{Fig:data_acquisition}. A high-speed voltage analog output PCI board (NI 6731, National Instruments) generates static or waveform voltages on two pairs of channels, each one commanding one dual-axis analog driver (Micromax 67320, Cambridge Technology) that controls one set of scanning mirrors. The TTL pulses generated by the two single photon counting modules are recorded by a PCI counting board (NI 6602, National Instruments). The synchronization between scanning and data acquisition is obtained by triggering the counters by a digital ``start'' signal generated by the analog output board when voltage generation starts. Since both PCI cards use direct memory access, the data are transferred through a buffer, ensuring high-speed operation. For FCS measurements, both APD signals are also sent to a multiple tau digital correlator (Flex02-12D, Correlator.com), that builds up in real time the temporal auto and cross correlations, and can optionally deliver photon counting histories for off-line software correlation analysis. The focus of the microscope stand is motorized.

The whole system is connected to a personal computer and controlled by a program developed in house in a LabVIEW (National Instruments) environment. It provides with a unique graphic user interface a control over the scanning parameters, the photon counts acquisition and processing, the correlator, and the microscope stand parameters (focus, ports, objective turret, filter turret, shutters, etc.).

\subsection{Sample preparation}

FCS measurements in solution were carried out using a Rhodamine 6G (diffusion coefficient $D = 280$~$\mu$m$^2$ at room temperature\cite{Magde1974}) solution, with concentrations ranging from 100~nM up to 1~$\mu$M. The excitation power, measured by a powermeter inserted before the sideport of the microscope stand (between L4 and M3, see Fig.~\ref{Fig:setup}) was 300~$\mu$W, a value that was checked to be low enough to prevent any effect of saturation or photobleaching.

Point spread function measurement were carried out using 100-nm diameter yellow-green fluorescent microspheres (FluoSpheres$^\circledR$ 505/515, Moleculer Probes), that have been dispersed on a cleaned microscope coverslip. Typical excitation power, measured as previously, was 5~$\mu$W.

Measurement on cells have been performed at room temperature on COS-7 cells, the GFP-tagged Thy1 protein of which has been transiently expressed using the protocol detailed in Ref.~\onlinecite{Lenne2006}.

\section{Theory}
\label{Sec:Theory}

\subsection{Point spread functions}

The response of a confocal fluorescence microscope is described by a 3D point spread function (PSF), that we denote  $\text{PSF}_\text{conf}$, which takes into account two contributions: i) the spatial distribution of the excitation intensity within the specimen, described by the excitation PSF, that we denote $\text{PSF}_\text{exc}$, ii) the collection efficiency, described by the intensity collection PSF, denoted $\text{PSF}_\text{coll}$, with\cite{Jonkman}
\begin{eqnarray*}
\text{PSF}_\text{conf}(\mathbf{r}) & = & \text{PSF}_\text{exc}(\mathbf{r}) \cdot \text{PSF}_\text{coll}(\mathbf{r})\\
&  = & (\text{PSF}_\text{exc} \cdot \text{PSF}_\text{coll})(\mathbf{r}).
\end{eqnarray*}
With these definitions, the intensity $I(\mathbf{r})$ recorded on an specimen $S(\mathbf{r})$, is given by
\begin{eqnarray}
I(\mathbf{r}) & = &  \int  S(\mathbf{R}) \cdot \text{PSF}_\text{conf}(\mathbf{r}-\mathbf{R}) \, \text{d}^3 \mathbf{R} \label{Eq:PSF}\\
 & = & (S \otimes \text{PSF}_\text{conf}) (\mathbf{r}) \nonumber \\
 & = & [S \otimes (\text{PSF}_\text{exc} \cdot \text{PSF}_\text{coll})] (\mathbf{r})\nonumber
\end{eqnarray}

\begin{figure}
\includegraphics{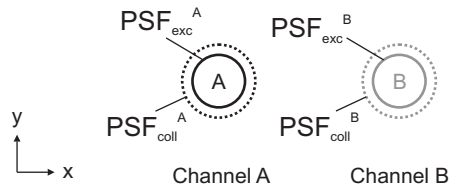}
\caption{\label{Fig:Volumes}Schematic view of the two measurement spots, with the corresponding lateral extents of the excitation (solid line) and collection (dashed line) PSFs. In this example, the two spots are addressing different locations.}
\end{figure}

Since our system is made of two independent channels, each of them will therefore be described by its own set of PSFs. A schematic view of the PSFs involved in our system is presented in Fig.~\ref{Fig:Volumes}, where only the spatial extent of the excitation and collection PSFs are represented in the $xy$ plane for a better clarity. This kind of representation constitutes a powerful way to figure out the respective roles of the two spots, their possible interaction, and more generally all degrees of freedom of the system. Note that the correct pinhole alignment performed on each channel before each campaign of measurement ensures that $\text{PSF}^\text{A}_\text{exc}$ and $\text{PSF}^\text{A}_\text{coll}$ (and the same for PSFs of channel B) are centered with respect to each other, as illustrated in Fig.~\ref{Fig:Volumes}, keeping in mind that these properties are conserved during the scanning due to the descanned detection scheme.

\subsection{Fluorescence correlation spectroscopy (FCS)}

The technique of FCS is based on the analysis of the fluctuation of the fluorescence intensity $F(t)$, by means of its temporal autocorrelation function $G(\tau)$, defined as
$$
G(\tau) = \frac{\langle F(t)F(t+\tau) \rangle}{\langle F \rangle^2},
$$
where the brackets $\langle \rangle$ indicate a temporal average. Since the PSF of the system can  be reasonably described by a Gaussian distribution\cite{Rigler1993}
\begin{equation}
\label{Eq:Gauss3D}
\text{PSF}_\text{conf}(\mathbf{r}) = I_0 \exp \left[
-2 \left( \frac{x^2}{\omega^2_x}+\frac{y^2}{\omega^2_y}+\frac{z^2}{\omega^2_{z}}\right) \right],
\end{equation}
it has been shown that the autocorrelation function for free diffusing fluorescent molecules can be written as\cite{Widengren1994}
\begin{eqnarray}
\label{Eq:FCS}
G(\tau) = 1 & + & \frac{1}{N} \left(1+\frac{4D\tau}{w^2_{xy}}\right)^{-1} \left(1+\frac{4D\tau}{w^2_{z}}\right)^{-1/2} \nonumber\\
& \times & \left[1+n_T \exp \left(- \frac{\tau}{\tau_T} \right) \right],
\end{eqnarray}
where $D$ is the translational diffusion coefficient, $N$ the average number of molecules in the sample volume, $n_T$ the triplet fraction and $\tau_T$ the triplet lifetime, and  $w_{xy} = w_x = w_y$. By fitting experimental data with the expression of Eq.~\ref{Eq:FCS}, it is possible either to measure the diffusion coefficient if $w_{xy}$ and $w_z$ are known, or to measure $w_{xy}$ and $w_z$ from a solution of known diffusion coefficient $D$.

\section{Calibrations}
\label{Sec:Calibrations}

\subsection{Distances within the sample plane}

The conversion factor between the command voltage and the transverse displacement of the spot in the sample has been calibrated for each of the two scanners by recording the image of calibrated micrometers (Graticules, Ltd). The total area that can be covered by scanning is limited in our microscope stand by the side port accessible diameter. With our objective, it is typically $200\times 200$~$\mu$m$^2$.

\subsection{Scanning speed}

\begin{figure}
\includegraphics{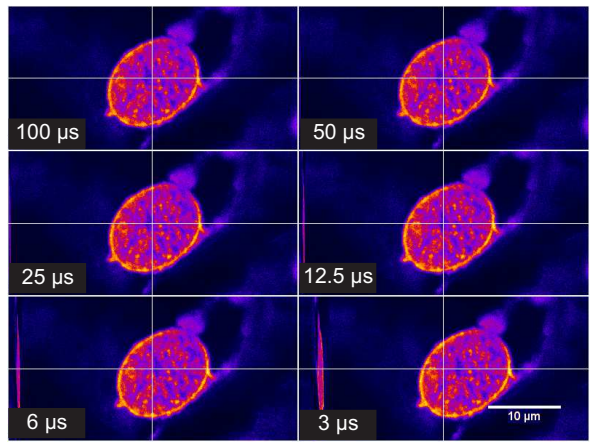}
\caption{\label{Fig:Shift} (Color online) Comparison of several images of a fixed sample acquired with different pixel dwell times (indicated in the corner of each image). In order to compare images of similar signal to noise ratio, an appropriate number of accumulations was set, keeping constant the total dwell time per pixel at a value of 100~$\mu$s (1 accumulation at 100~$\mu$s, 2 accumulations for 50~$\mu$s, etc.). For the shortest dwell times, the image shift is clearly visible. In addition the narrow left part of the image shows the signal that has been recorded when the spot on its way back, between successive line scans.}
\end{figure}

Since the main purpose of the system is FCS measurements, highly sensitive detectors have been chosen, which operate in photon counting mode. The count rate is therefore limited to a few $10^6$ counts per second, i.e., a few counts per microsecond. Thus, as far as imaging is concerned, the main limitation in terms of acquisition rate is not the velocity of the scanning system itself (that can reach dwell time down to the micro second), but the dynamics of photon counts within the image, that need a sufficient dwell time, and should provide a contrast with an acceptable signal to noise ratio. In practice, the minimum pixel dwell time is typically $100$~$\mu$s, a value that can be obtained either by one single scan, or by the accumulation of several frames.

Images are obtained by raster scanning, i. e., all lines from an image are recorded sequentially, the spot being scanned for each line in the same direction. For extremely fast scanning conditions (dwell time smaller than 50~$\mu$s), the mechanical inertia of the scanners causes the beam displacement to be delayed with respect to the signal acquisition. In these conditions, photon counts that are attributed to one pixel actually may actually come from an earlier location, and the final image appears therefore slightly shifted along one direction, as illustrated in Fig.~\ref{Fig:Shift}. Fortunately, no deformation has been noticed, so that distance measurements are unaffected by this effect.

\subsection{Transverse channel matching}

\begin{figure}
\includegraphics{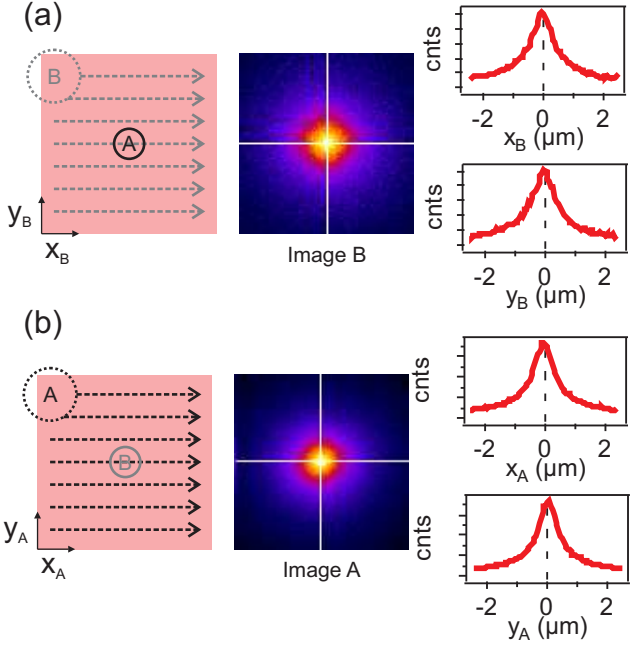}
\caption{\label{Fig:crossimaging} (Color online) Calibration of channel matching by cross imaging in a solution of Rhodamine 6G. (a) Left, schematic of the mode of operation, with PSF extents as defined in Fig.~\ref{Fig:Volumes}; center, image recorded on channel B; right, cross section of the image along axis plotted in white. Left axis is in arbitrary unit, minimum is zero. (b) Same procedure, vice versa.}
\end{figure}

Since the two spots are independently controlled by their own scanners, their location is given in their own set of coordinates. For all measurements that will involve two spots, it is crucial that these two coordinate systems match perfectly. A first coarse alignment is performed by placing the two scanning systems symmetrically with respect to the beam splitter (denoted BS in Fig.~\ref{Fig:setup}). Then, a finer calibration is realized by cross imaging, using a homogeneous solution of Rhodamine 6G (1~$\mu$M, typically). Spot A is on, static, located for instance at $(x_\text{A} = 0, y_\text{A} = 0)$ while channel B, with laser excitation off, is performing a raster scan. The maximum of fluorescence is collected by channel B when the centers of $\text{PSF}^\text{B}_\text{coll}$ and $\text{PSF}^\text{A}_\text{exc}$ match, which allows to measure the location of spot A in the system of coordinates of B. By tilting scanner B, it is possible to move spot A to $(x_\text{B} = 0, y_\text{B} = 0)$, and therefore to ensure that $(x_\text{A} = x_\text{B}, y_\text{A}= y_\text{B})$, as illustrated in Fig.~\ref{Fig:crossimaging}a, with a typical accuracy of 20~nm. Without any further alinement, a quick check can be realized by operating vice versa, as illustrated in Fig.~\ref{Fig:crossimaging}b. Finally, it was check that the channel matching occurs for all spot locations within the scanning field.

\subsection{Axial distance between spots}
\label{Sec:zscan}

\begin{figure}
\includegraphics{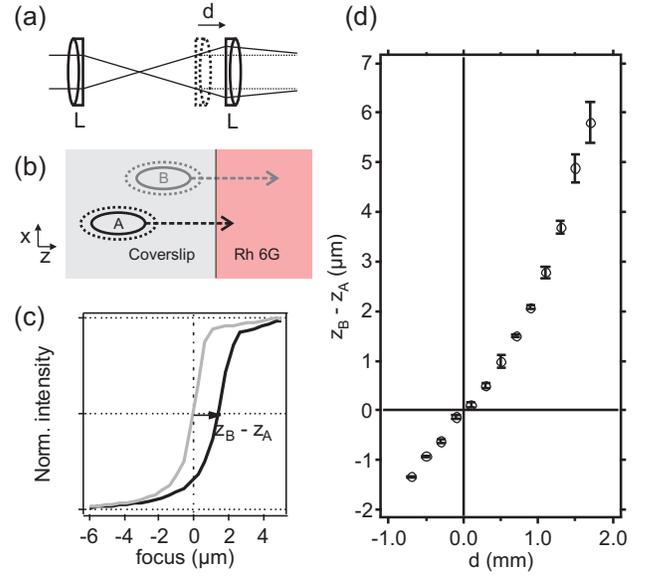}
\caption{\label{Fig:zscan} a) Schematic of the telescope T inserted in channel A and allowing to modify the plane of focusing of spot A. The distance $d$ is defined as the shift to the nominal distance between the two lenses. b) Schematic view of the spots focused in different planes in the case $d>0$ and laterally shifted. c) Example of intensity profiles recorded by z-scan. d) Values of $z_B-z_A$ measured versus $d$.}
\end{figure}

As discussed in section~\ref{Sec:Setup}, the scanning plane of spot A can be moved backward or forward with respect to spot~B by simply acting on a telescope (denoted T in Fig.~\ref{Fig:setup}). Figure~\ref{Fig:zscan}a illustrates the telescope system, where the shift to the nominal position is denoted~$d$. The resulting axial distance can be monitored by performing a scan with the two spots through the interface between a coverslip and a fluorescent solution, as illustrated in Fig.~\ref{Fig:zscan}b. Note that the two spots have been laterally split apart by 5~$\mu$m in order to prevent from unwanted crosstalk. The signal recorded on both channels shows the same typical smooth step shape represented in Fig.~\ref{Fig:zscan}c, where the half value of maximum intensity is obtained when center of the spot is exactly at the interface. The axial distance $z_B - z_A$ can be therefore measured as the distance between the two half maxima, with an accuracy of 100~nm. The relationship between $z_B - z_A$ and $d$ is plotted in Fig.~\ref{Fig:zscan}d. It shows that an axial shift up to $6$~$\mu$m can be reached with this system. The issue of the spot quality in off-plane configuration will be addressed in section~\ref{Sec:offplane}.

\subsection{Direct independent measurement of excitation and collection PSFs}

\begin{figure}
\includegraphics{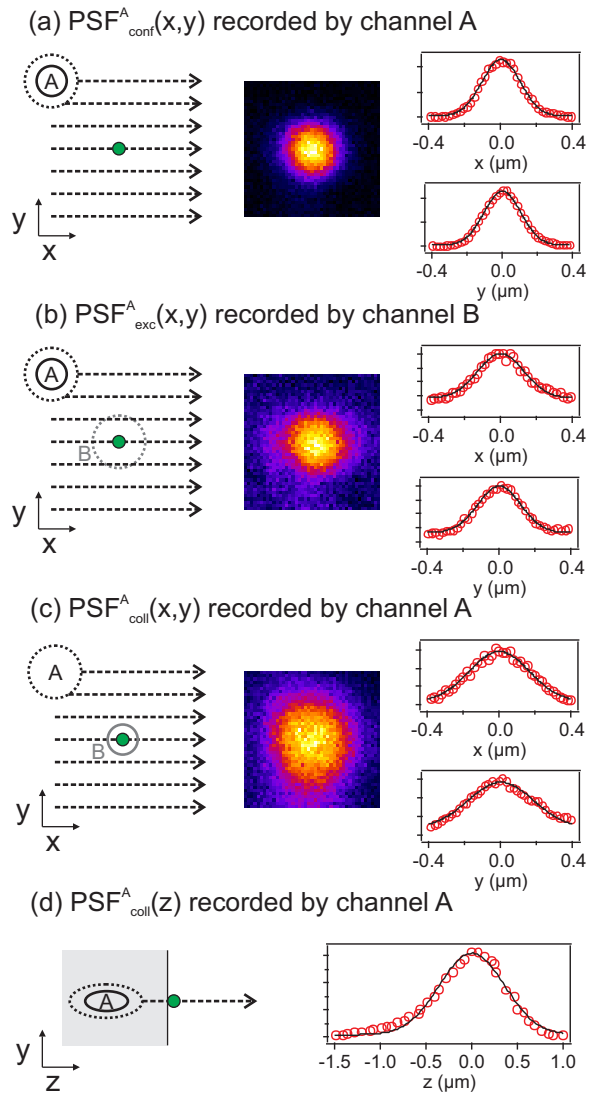}
\caption{\label{Fig:PSFs} (Color online) (a-c) Left: Measurement protocole, the green circle is the fluorescent microsphere. Center: Recorded intensity distribution. Right: Cross section (circles) and best fit (solid line)  according to Eq.~\ref{Eq:Gauss3D}. (d) Measurement protocole for $\text{PSF}^\mathrm{A}_\text{conf}(z)$ and resulting intensity distribution, with best fit.}
\end{figure}

The spatial resolution of a confocal microscope system is usually assessed by recording the image of a subwavelength isolated fluorescent microsphere, that can therefore be considered as a point source. By replacing the specimen function $S$ in Eq.~\ref{Eq:PSF} by a Dirac distribution $\delta$, the intensity recorded by channel A is indeed given by
$$
I^\mathrm{A}(\mathbf{r}_\mathrm{A}) = \int  \delta(\mathbf{R}) \cdot \text{PSF}^\mathrm{A}_\text{conf}(\mathbf{r}_\mathrm{A}-\mathbf{R}) \, \text{d}^3 \mathbf{R} = \text{PSF}^\mathrm{A}_\text{conf}(\mathbf{r}_\mathrm{A}).
$$
The shape of the intensity distribution is a good indicator of the overall alinement of channel A, while its spatial extends $w_{xy}$ and $w_{z}$ are directly related to the spatial resolution. A sketch of the measurement procedure, an example of recorded intensity distribution, and the corresponding cross-section are represented in Fig.~\ref{Fig:PSFs}a. In addition, during this measurement, if the collection volume of channel B is overlapping the microsphere, with laser off (see sketch of Fig.~\ref{Fig:PSFs}b), the intensity that is recorded by channel B while A is scanning (laser on) can be written as
\begin{eqnarray*}
I^\mathrm{B}(\mathbf{r}_\mathrm{A}) &=& \int  \delta(\mathbf{R}) \cdot \text{PSF}^\mathrm{A}_\text{exc}(\mathbf{r}_\mathrm{A}-\mathbf{R}) \cdot \text{PSF}^\mathrm{B}_\text{coll}(\mathbf{r}_0)\, \text{d}^3 \mathbf{R} \\
&=& \text{PSF}^\mathrm{B}_\text{coll}(\mathbf{r}_0) \cdot \text{PSF}^\mathrm{A}_\text{exc}(\mathbf{r}_\mathrm{A})\\
&\propto &\text{PSF}^\mathrm{A}_\text{exc}(\mathbf{r}_\mathrm{A}).
\end{eqnarray*}
The intensity recorded by channel B is therefore proportional $\text{PSF}^\mathrm{A}_\text{exc}$. Finally, if the excitation is delivered to the sphere by spot B (static, laser on), scanning with channel A (laser off, see sketch of Fig.~\ref{Fig:PSFs}c) will provide an intensity distribution proportional to $\text{PSF}^\mathrm{A}_\text{coll}(\mathbf{r}_\mathrm{A})$.

Since the measurement schemes of Figs.~\ref{Fig:PSFs}a and \ref{Fig:PSFs}b can be performed simultaneously, only two measurements (b and c) are needed to provide independently all three PSFs involved in one channel. The characterization of the second channel will be performed in an identical way by reversing the role played by the two channels. Figure~\ref{Fig:PSFs} summarized the protocols and results obtained for channel A, intensity cross sections along $x$ and $y$ axis, as well as the best fit (solid line) using Eq.~\ref{Eq:Gauss3D}. Although this later assumption of a 3D Gaussian distribution is only an approximation in the case of confocal PSFs,\cite{Rigler1993} it describes reasonably the present data, especially in the transverse direction. The corresponding widths for all PSFs for channel A and B (raw data not shown) are summarized in Table~\ref{tab:widths}. These values show that the two channels possess very similar features, close to the diffraction limit of the microscope objective.

\begin{table}
\caption{\label{tab:widths}Summary of all PSF widths measured for both channels, as defined by Eq.\ref{Eq:PSF}. Values are given in nm.}
\begin{ruledtabular}
\begin{tabular}{lcccccc}
 & \multicolumn{3}{c} {Channel A} & \multicolumn{3}{c} {Channel B} \\
 &$w_x$ &   $w_y$   &   $w_z$   & $w_x$   &   $w_y$   & $w_z$ \\
\hline
$\text{PSF}_\text{exc}$  &   256 & 235 & - & 267 & 250 & - \\
$\text{PSF}_\text{coll}$ &  345 & 392 & - & 354 & 351 & - \\
$\text{PSF}_\text{conf}$ & 211 & 206 & 709 & 223 & 219 & 788 \\
$\text{PSF}_\text{conf}$ calculated\footnotemark[1] & 206 & 202 & & 213 & 204 & \\
$\text{PSF}_\text{conf}$ (FCS) & \multicolumn{2}{c} {180} & 902 & \multicolumn{2}{c} {190} & 952 \\
\end{tabular}
\end{ruledtabular}
\footnotetext[1]{Using the widths of $\text{PSF}_\text{exc}$ and $\text{PSF}_\text{coll}$ and the property that the product of two Gaussian distributions of widths $w_1$ and $w_2$ is Gaussian, with $w = \frac{w_1 w_2}{\sqrt{w_1^2+w_2^2}}$.}
\end{table}

\begin{figure}
\includegraphics{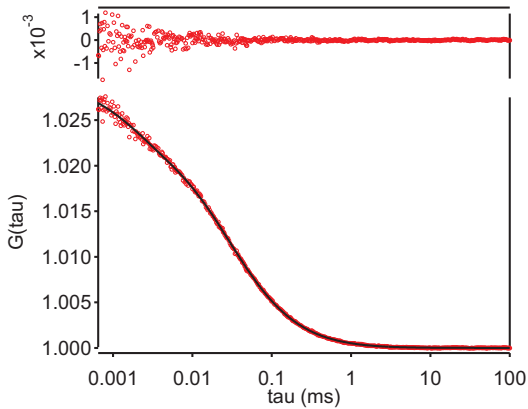}
\caption{\label{Fig:FCS_sol} (Color online) Example of FCS measurement by channel A performed in a solution of Rhodamine 6G. Autocorrelation data have been averaged over 10 measurements of duration 10~s. Data are plotted with circles, while the best fit using Eq.~\ref{Eq:FCS} is plotted with a solid line. The residues are plotted in the top graph. The corresponding fitting parameters are $N = 42$, $w_{xy} = 180$~nm, $w_z = 902$~nm, $n_T = 0.24$, and $\tau_T=1.6$~$\mu$s.}
\end{figure}

These values have been compared to those obtained by FCS performed in a solution of Rhodamine 6G, as described in the experimental section. The obtained autocorrelation function for channel A is plotted in Fig.~\ref{Fig:FCS_sol}, as well as the best fit using Eq.~\ref{Eq:FCS}. The corresponding values of $w_{xy}$ for both channels have been reported in Table~\ref{tab:widths}. The obtained values of the spatial extend are in the same range as the ones obtained by imaging. In spite of the slight difference that could be explained by the uncertainty on the diffusion coefficient of the solution, FCS measurement remains a valid method to assess the overall size of the confocal volume. Moreover, unlike single micro-sphere imaging, this method can be performed in a few seconds, and can be easily automated, as it will be illustrated below.

\subsection{Off-axis PSF}

\begin{figure}
\includegraphics{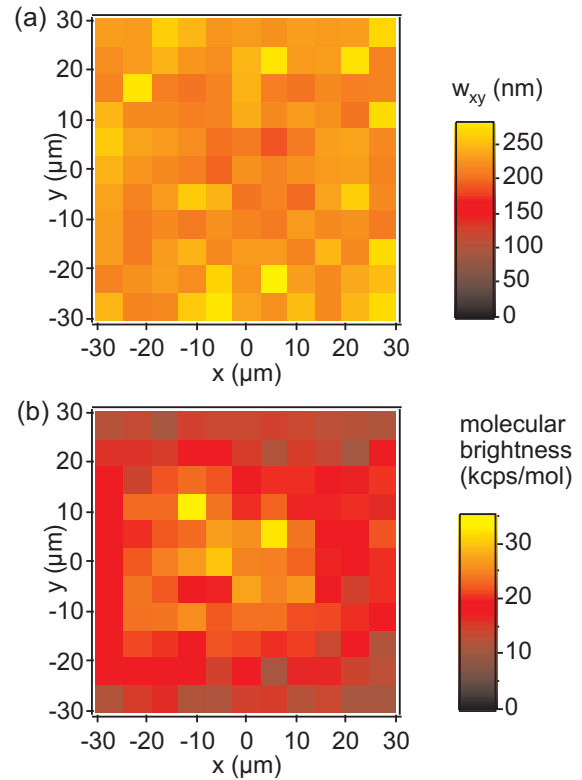}
\caption{\label{Fig:mapFCS} (Color online) Colormaps showing how the performances of the confocal system depend on the spot location within the scanning range. (a) Map of the values of the lateral width $w^\mathrm{A}_{xy}$ of $\text{PSF}_\mathrm{conf}^\mathrm{A}$. (b) Map of the molecular brightness for channel A.}
\end{figure}

In order that the system performs FCS measurement at arbitrary locations, it is important that the shape and size of the confocal volume are not affected when the spot is focused out of the optical axis of the objective. The measurement of the confocal volume using FCS analysis of a solution of Rhodamine 6G reported above has been extended over a large number of discrete points within a scanning range of 60$\times$60~$\mu$m$^2$ around the center. Autocorrelation functions recorded at each point were fitted according to Eq.~\ref{Eq:FCS}. The corresponding values of lateral extend $w_{xy}$ are plotted as a color map in Fig.~\ref{Fig:mapFCS}a. They show a slight variation over the scanning range, the extreme values being found only far from the center. A more pronounced behavior was observed for the average number of molecule $N$ (data not shown). The resulting map of molecular brightness, given by the ratio of the intensity to $N$, is reported in Fig.~\ref{Fig:mapFCS}b. The drop by a factor of~3 at the corners of the scanning field can be explained by the increased value of $N$ and to a lesser extent by the degradation of the microscope objective transmission in off-axis conditions.

Measurements performed on channel~B (data not shown) show a similar behavior. This confirms the ability of the system to carry out dual spot measurements at the spatial scale of a cell.

\subsection{Off-plane PSF}
\label{Sec:offplane}

\begin{figure}
\includegraphics{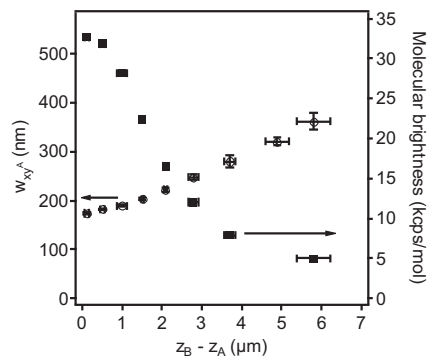}
\caption{\label{Fig:Zeffect}Plot of dependence of lateral width $w_{xy}^\mathrm{A}$ (left axis) and molecular brightness (right axis) when spot A is moved axially with respect to the nominal focal plane.}
\end{figure}

The same strategy was used to characterize the confocal volume of channel A when its plane of focus is changed by acting on the variable telescope. The values of the lateral width $w^\mathrm{A}_{xy}$ and of the molecular brightness are plotted for different axial distances between spots in Fig.~\ref{Fig:Zeffect}. Note that the measurements performed for $z_\mathrm{B}-z_\mathrm{A} < 0$, i.~e., spot B being the closest to the microscope objective, gave rise to data that cannot be fitted properly. This was also the case to a lesser extent for $z_\mathrm{B}-z_\mathrm{A} > 4$~$\mu$m. This is probably due to a strong deformation of the spot, with makes the assumption of a three-dimensional gaussian shape no longer valid.

\section{Examples of measurements}
\label{Sec:Examples}

The main original feature of this system is its fully independent dual spot nature. Although the presented configuration includes photon counting and temporal correlation analysis, it can of course be extended by implementing dedicated modules, such as lifetime analysis, polarization-resolved excitation and/or collection, etc.\cite{Ferrand2008} However, care has to be taken in order to prevent from unwanted cross-talk between channels. Indeed, as it was extensively exploited in section~\ref{Sec:Calibrations}, the two channels can interact in case of spatial overlap, because no discrimination was possible using two channel of identical spectral features under a continuous wave excitation. In case spatial overlap cannot be avoided, an additional discrimination scheme should be implemented.

Although biophysical investigations are much beyond the scope of this instrumental article, a couple of measurements are described in this section. They have been performed on living cells and illustrate well the versatility of the setup.

\subsection{Simultaneous imaging}

\begin{figure}
\includegraphics{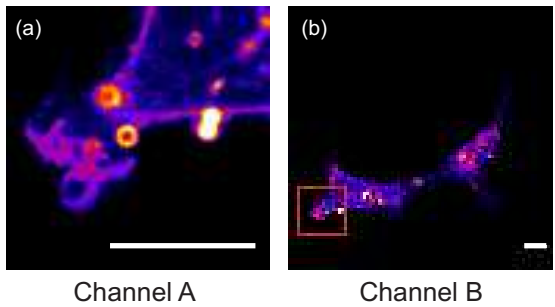}
\caption{\label{Fig:Simultaneous_imaging} (Color online) Example of simultaneous imaging. These two images have been acquired simultaneously on the COS7 (Thy1-GFP) living cell with the same frame rate, channel A (a) offering a high definition imaging of a limited area (a), while channel B provides a simultaneous low-definition overview of the sample (b). The red square on image B indicates the area scanned by channel A. Scale bars are 10~$\mu$m.}
\end{figure}

Because laser scanning confocal imaging relies on a sequential acquisition scheme, high definition imaging with a high repetition rate can only be performed on a restricted observation area, eliminating therefore any possibility of control of the sample in its entire scale. Because the two scanning channels of our system are independent, two confocal acquisition can be performed simultaneously at two different locations in a sample, with identical or different magnification. As illustrated in Fig.~\ref{Fig:Simultaneous_imaging}, one small area on a sample can be imaged with relatively high definition and frame rate, while the other channel allow to image simultaneously the entire cell (at the same frame rate, but a lower definition), enabling therefore to control the entire cell, and to monitor any drift or perturbing event.

\subsection{Simultaneous imaging and FCS analysis}

\begin{figure}
\includegraphics{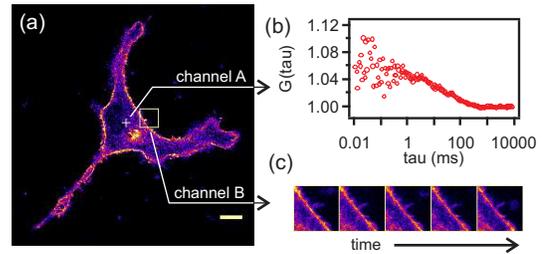}
\caption{\label{Fig:FCS_imaging} (Color online) a) Overview of the COS7 (Thy1-GFP) cell recorded by confocal imaging. Scale bar is 10~$\mu$m. The FCS measurement was then performed by channel A at the location indicated by a cross, while a sequence of control images was recorded by channel B at the location indicated by a square. b) Autocorrelation curve recorded by channel A. c) Extract of the sequence of images recorded by channel~B.}
\end{figure}

The issue of the overall control of the sample is even more relevant in the case of a FCS experiment. Indeed, since this later technique relies on the hypothesis of stationarity, any perturbation such as mechanical drift, or passage of unwanted aggregates, may disturb the acquisition and produce erroneous measurements. Figure~\ref{Fig:FCS_imaging} is an example of measurement performed on a living cell. First, an image of the cell is recorded (Fig.~\ref{Fig:FCS_imaging}a). Then, channel A is dedicated to a static FCS measurement (Fig.~\ref{Fig:FCS_imaging}b), while channel B performs simultaneously a continuous image acquisition on an area slightly apart (Fig.~\ref{Fig:FCS_imaging}c). Therefore, FCS data are supported by the additional information of the image sequence that can help to select measurements that are in good agreement with the FCS hypotheses.

\subsection{Dual spot scanning FCS}

\begin{figure}
\includegraphics{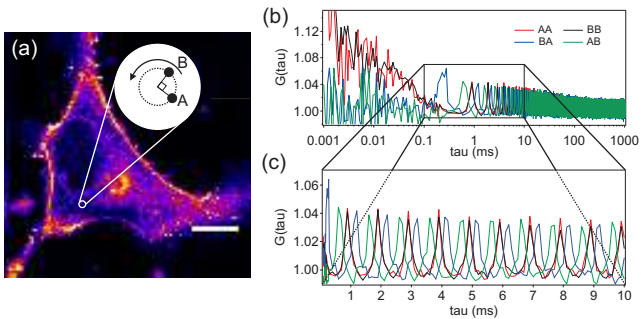}
\caption{\label{Fig:sFCS} (Color online) a) Overview of the COS7 (Thy1-GFP) cell under study. Scale bar is 10~$\mu$m. The scanning orbit is indicated by a circle. b) Correlation data. Correlation $G_{AB} = \frac{\langle I_A(t)I_B(t+\tau) \rangle}{\langle I_A \rangle \langle I_B \rangle}$ is denoted $AB$, and so on. Note that correlation values for $\tau < 0.3$~ms are given by the hardware correlator, while the ones for larger delays have been software computed off-line. c) Same as (b), but plotted within a restricted range in a linear timescale.}
\end{figure}

Finally, a measurement protocol involving on a genuine dual spot analysis is presented in Fig.~\ref{Fig:sFCS}. As illustrated in the inset of Fig.~\ref{Fig:sFCS}a, the two measurement volumes are scanned in a periodic fashion with the same angular velocity on the same circular orbit located on the cell membrane, spot A following B with a delay of a quarter of an orbit. An orbit radius of 0.5~$\mu$m was chosen, as well as a rotation frequency of 1~kHz. The four temporal correlations ($G_{AA}$, $G_{BB}$, $G_{AB}$, and $G_{BA}$) are plotted in Figs.~\ref{Fig:sFCS}b and \ref{Fig:sFCS}c.

As it has been detailed in the literature,\cite{Berland1996} the temporal correlations measured in scanning FCS present, in addition to the usual decay due to translational diffusion, a modulation with a period given by the scanning frequency, i.~e., 1~ms in the present case. The peaks of correlation are obtained for delays that bring to correspondence photon counts acquired at the same point of the orbit. For the two autocorrelations $G_{AA}$ and $G_{BB}$, this happens for $\tau = 1$~ms, 2~ms, 3~ms, and so on, as it is clearly visible in Fig.~\ref{Fig:sFCS}c. Because spot $A$ is following spot $B$, the first peak of $G_{BA}$ corresponds to a quarter of an orbit, i.~e., is obtained at $\tau = 0.25$~ms, the following peaks occurring at $\tau = 1.25$~ms, 2.25~ms, etc. Finally, peaks for correlation $G_{AB}$ are measured at $\tau = 0.75$~ms, 1.75~ms, etc.

Obtaining scanning FCS data for short delays is a challenging issue in a single spot geometry because the shortest delay is directly given by the fastest scanning period, which is usually limited by mechanical response of the scanning system.\cite{Petrasek2008b} Without modifying the scanning features, the dual spot scheme that we propose allows to address correlation delays that are significantly shorter. Therefore we believe that this approach can provide more accurate measurements. In the present example, although the scanning period is limited to 1~ms, scanning FCS cross correlation data have been obtained for delays down to 0.25~ms. Given a scanning period, the limitation for the short delays is now constituted by the crosstalk between the two channels. The full potential of dual spot scanning FCS will be addressed in a dedicated article.

\section{Conclusion}

A versatile dual spot FCS system has been presented. A complete calibration protocol has been detailed, which allow to control accurately the location, size and shape of the two measurement volumes. A method for separating the contribution of excitation and collection volume has been proposed. The spot quality appeared to be compatible with FCS measurements in a wide transverse area, and in a few microns apart the nominal focus plane. The versatility of the setup was illustrated by measurements carried out on living cells using two spots in scanning and/or static measurement modes.

\begin{acknowledgements}
This project was funded by the French Agence Nationale de la Recherche under contract ANR-05-BLAN-0337-02 and by R\'egion Provence Alpes C\^ote d'Azur. The authors are grateful to Emmanuel Schaub for his advice for setting up the scanning system.
\end{acknowledgements}


\end{document}